\documentclass[aps,prl,showpacs,twocolumn,preprintnumbers,superscriptaddress]{revtex4-1}
\usepackage{amsmath,amssymb}%
\usepackage{graphicx}
\usepackage{dcolumn}
\usepackage{bm}

\begin{document}
\title{Scrutinizing Hall effect in Mn$_{1-x}$Fe$_{x}$Si: Fermi surface evolution and hidden quantum criticality}
\date{\today}

\author{V.~{}V.~{}Glushkov}
\email{glushkov@lt.gpi.ru}
\author{I.~{}I.~{}Lobanova}
\affiliation {Prokhorov General Physics Institute of RAS, 38 Vavilov str.,  119991 Moscow, Russia} 
\affiliation{Moscow Institute of Physics and Technology , 9 Institutskiy lane, 141700 Dolgoprudny, Moscow region, Russia}
\author{V.~{}Yu.~{}Ivanov}
\author{V.~{}V.~{}Voronov}
\affiliation {Prokhorov General Physics Institute of RAS, 38 Vavilov str.,  119991 Moscow, Russia} 
\author{V.~{}A.~{}Dyadkin}
\affiliation{Swiss-Norwegian Beamlines at the European Synchrotron Radiation Facility, 38000 Grenoble, France}
\author{N.~{}M.~{}Chubova}
\author{S.~{}V.~{}Grigoriev}
\affiliation{Petersburg Nuclear Physics Institute, Gatchina, 188300 Saint-Petersburg, Russia}
\author{S.~{}V.~{}Demishev}
\affiliation {Prokhorov General Physics Institute of RAS, 38 Vavilov str.,  119991 Moscow, Russia} 
\affiliation{Moscow Institute of Physics and Technology , 9 Institutskiy lane, 141700 Dolgoprudny, Moscow region, Russia}

\begin{abstract}
 
 Separating between ordinary (OHE) and anomalous (AHE) Hall effect in the paramagnetic phase of Mn$_{1-x}$Fe$_{x}$Si reveals OHE sign inversion associated with the hidden quantum critical (QC) point $x^*\sim0.11$. The semimetallic behavior at intermediate Fe content leads to verifiable predictions in the field of fermiology, magnetic interactions and QC in Mn$_{1-x}$Fe$_{x}$Si. The change of electron and hole concentrations is considered as a  \textquotedblleft driving force\textquotedblright for tuning the QC regime in Mn$_{1-x}$Fe$_{x}$Si via modifying of RKKY exchange interaction within the Heisenberg model of magnetism.
 
 \end{abstract}

\pacs{72.15.Gd; 75.30.-m; 75.30.Kz}

\maketitle

Studying of the ordinary Hall effect (OHE) in quantum critical (QC) regime is an important tool, which allows choosing between various scenarios of non-Fermi liquid behavior in various strongly correlated electron systems \cite{Paschen04,Combier13,Friedemann10}. In the case of localized magnetic moments (LMM) a collapse of the Fermi surface (FS) expected exactly at the quantum critical point (QCP) results in an abrupt change of the Hall constant at zero temperature \cite{Paschen04,Friedemann10}. In contrast, no direct evidence of the Lifshitz transition at QCP  \cite{Combier13} is provided for itinerant magnets in the spin density wave model of quantum criticality \cite{Friedemann10}.  

\begin{figure}[t]
	\includegraphics[width=0.8\linewidth]{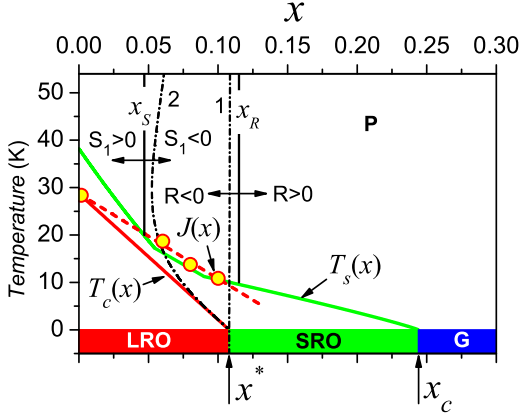}
	\caption{\label{fig1}(Colour on-line) Mn$_{1-x}$Fe$_x$Si magnetic phase diagram. P phase corresponds to $T>T_s(x)$, $T_s(x)$ is the onset of SRO. LRO phase boundary $T_c(x)$  \cite{Grigoriev11,Demishev13} is accompanied by the $T_c(0)J(x)/J(0)$ data plotted from experimental values of exchange energy \cite{Grigoriev09}. Solid lines at $x_S$ and $x_R$ (P phase) separate the various regimes of AHE and OHE (see text for details). Dash-dotted lines point to (1) the hidden QCP $x^*\sim0.12$ and (2) the  crossover between classic and quantum fluctuations \cite{Demishev13}}
\end{figure}

This apparent distinction between localized and itinerant behaviour stimulates a particular interest to the study of OHE in Mn$_{1-x}$Fe$_{x}$Si solid solutions. Recently, comprehensive neutron scattering study  \cite{Grigoriev09,Grigoriev11} together with magnetic data  \cite{Nishihara84,Bauer10,Demishev13} and specific heat measurements  \cite{Bauer10} discovered a QCP  corresponding to the suppression of spiral phase with long-range magnetic order (LRO) in  Mn$_{1-x}$Fe$_{x}$Si. This QCP located at $x^*\sim0.11-0.12$ (Fig.~{}\ref{fig1})  \cite{Grigoriev09,Grigoriev11,Nishihara84,Bauer10,Demishev13} is hidden by a surrounding phase with short-range magnetic order (SRO)  \cite{Grigoriev11,Bauer10,Demishev13} that agrees well with the theoretical models  \cite{Tewari06,Kruger12}. This SRO phase referred sometimes as chiral spin liquid \cite{Tewari06} is destroyed at the second QCP $x_c\sim0.24$ (Fig.~{}\ref{fig1}) \cite{Demishev13}. Diverging of magnetic susceptibility $\chi(T)\sim1/T^\xi (\xi=0.5-0.6)$ at $x>x_c$  \cite{Tewari06} is proved to be a fingerprint of disorder-driven Griffiths (G) phase consisting of separated spin clusters  \cite{Bray87,Griffiths69}. So not only modulation of exchange interactions, which seems to induce the first QCP at $x^*$  \cite{Grigoriev11,Demishev13}, but also strong disorder effects  impact significantly on quantum criticality in Mn$_{1-x}$Fe$_x$Si. 

However, some essential features of QC behavior in this system have not been recognized up to now. Firstly, the critical temperature of the LRO phase $T_c(x)$ does not follow the $x$ dependence of the ferromagnetic (F) exchange $J(x)$, which turns zero for Fe content exceeding $x^*$ (Fig.~{}\ref{fig1}) \cite{Grigoriev09}. As long as LRO in Mn$_{1-x}$Fe$_x$Si originates from  competing F exchange and Dzyaloshinskii-Moriya  interaction (DMI) \cite{Grigoriev09,Grigoriev11} and DMI does not result in any magnetic ordering itself, the different behavior of $T_c(x)$ and $J(x)$ indicates an intrinsic complexity of magnetic interactions in Mn$_{1-x}$Fe$_x$Si. Secondly, although the fragmentation into spin clusters expected at $x_c$ is supported by the theoretical model  \cite{Demishev13} taking this experimental fact for granted, the formal percolation in Mn magnetic ions at $x_c\sim0.24$ is not broken. So the factors driving the change in the topology of magnetic subsystem need to be clarified. Thirdly, despite the fact that the Mn$_{1-x}$Fe$_x$Si solid solutions are often considered as itinerant magnets \cite{Tewari06,Kruger12}, LDA calculations  \cite{Corti07} and recent magnetic resonance and magnetoresitance studies  \cite{Demishev12,Demishev14} favor the alternative explanation based on Heisenberg LMM of Mn ions. Therefore, different behavior of the Hall effect in QC systems with LMM and itinerant magnets \cite{Paschen04,Combier13,Friedemann10} makes it possible to shed more light on the microscopic mechanisms of quantum criticality in Mn$_{1-x}$Fe$_x$Si.

This Letter addresses the aforementioned problems through the study of the Hall effect in the paramagnetic (P) phase of Mn$_{1-x}$Fe$_x$Si (Fig.~{}\ref{fig1}). Currently, a noticeable discrepancy (about two times) in the electron concentration is reported even for pure MnSi  \cite{Lee07,Neubauer09}. Moreover, the recent study of the Hall effect in Mn$_{1-x}$Fe$_x$Si has initiated a pessimistic conclusion that any correct Hall constant can hardly be estimated because OHE in this system is much less than anomalous Hall effect (AHE)  \cite{Franz14}. This difficulty is shown to be overcome by the implementation of the data analysis developed recently for MnSi \cite{Glushkov15} allowing the reliable determination of the OHE and AHE contributions from experiment.

\begin{figure}[b]
\includegraphics[width=0.85\linewidth]{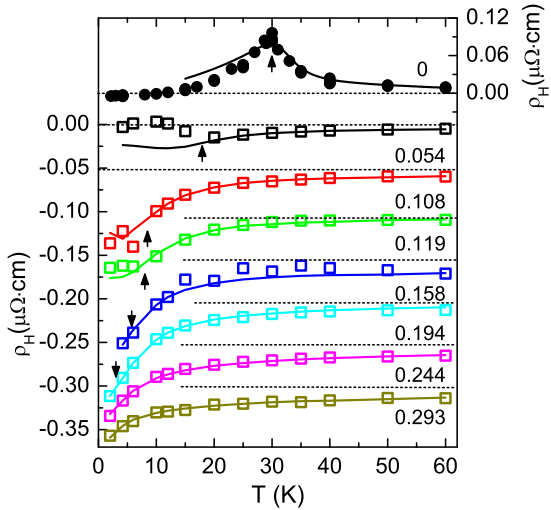}
\caption{\label{fig3}(Colour on-line) Hall resistivity $\rho_H(T,B_0)$ for pure MnSi (circles, $B_0=0.31$~T) and Mn$_{1-x}$Fe$_x$Si (squares, $B_0=0.5$~T). Arrows indicate the onset of SRO at $T_s(x)$. Solid lines are the fits by $\rho_H=R_HB_0+\mu_0S_1\rho{M}$ with $R_H(x)$ and $S_1(x)$ shown in Fig.~{}\ref{fig5},a. For clarity the $\rho_H(x>0.1)$ data are shifted down by $50~n\Omega\cdot$cm.}
\end{figure}

Experimental details and the set of Hall resistivity $\rho_H(B,T)$ and magnetization $M(B,T)$ data are resumed in the supplement \cite{Sup1}. Low field Hall resistivity extracted from the $\rho_H(B,T_0)$ data  for $B_0=0.5~T$ (Fig.~{}\ref{fig3}) shows clearly that $\rho_H$ decreases with lowering of temperature in the P-phase for all studied crystals. Distinct anomalies in the $\rho_H(T)$ data are identified at the transition into the SRO phase (arrows in Fig.~{}\ref{fig3}). The most prominent downturn of low temperature Hall resistivity occurs for  $x=0.194$ close to the SRO phase boundary at the second QCP $x_c\sim0.24$ (Fig.~{}\ref{fig3}).

The strong $\rho_H(T)$ dependences (Fig.~{}\ref{fig3}) allow applying the procedure  developed earlier to separate between OHE and AHE in the P phase of MnSi \cite{Glushkov15}. The functional form of the anomalous term $\rho_H^a=\mu_0S_n\rho^nM$ in $\rho_H=R_HB+\rho_H^a$ depends on the scattering mechanism of charge carriers \cite{Nagaosa10}. For Mn$_{1-x}$Fe$_x$Si the exponent $n=2$ corresponds to the intrinsic AHE related to $k$-space Berry phase effects \cite{Franz14}. The same value of $n=2$ was also predicted for extrinsic side-jump scattering \cite{Nagaosa10}. However, studies of Mn$_{1-x}$Fe$_x$Si single crystals  \cite{Franz14} and epitaxial MnSi thin films  \cite{Li13} prove evidently that the sample-dependent contribution AHE from side-jump scattering can be neglected. The case of $n=1$ is associated with skew scattering on LMM due to spin-orbit coupling  \cite{Nagaosa10}.

\begin{figure}[t]
	\includegraphics[width=0.85\linewidth]{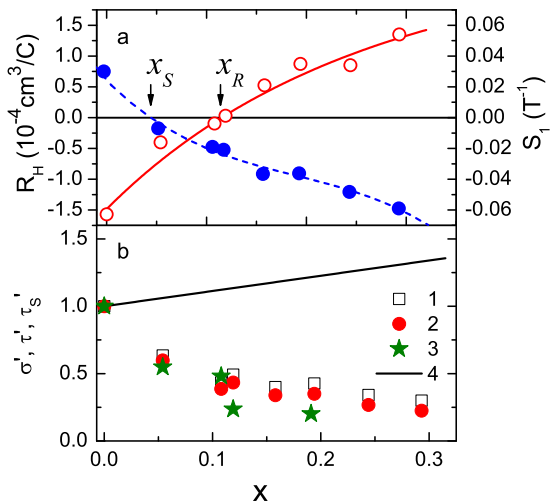}
	\caption{\label{fig5}(Colour on-line) a) Hall constant $R_H$ (open circles) and AHE coefficient $S_1$ (filled circles) in Mn$_{1-x}$Fe$_x$Si. Solid line is the fit by Eq.(1). (b) Reduced conductivity $\sigma(x)/\sigma(0)$ (1), electron transport relaxation time $\tau(x)/\tau(0)$ (2) and spin relaxation time $\tau_S(x)/\tau_S(0)$   \cite{Demishev11,Demishev14} (3) for $T=$30 K. Line 4 is the $\sigma(x)/\sigma(0)$ dependence expected for $\tau_e(x)=$const. }
\end{figure}

Comparison of the $\rho_H(T,B_0)$ data within these AHE scenarios shows that the best linear fits of the data in the P-phase are obtained in the $\rho_H/B_0=f(\rho{M}/B_0)$ plot \cite{Sup1}. This finding is supported by excellent agreement between the experimental $\rho_H(B,T)$ data and the curves calculated from the resistivity and magnetization in the $\rho_H^a\propto\rho{M}$  approximation (Fig.~{}\ref{fig3}). The $R_H(x)$ and $S_1(x)$ data extracted from the linear fits (Fig.~S4 \cite{Sup1}) are summarized in Fig.~{}\ref{fig5}a. Rising of Fe content results in the opposite trends of OHE and AHE changing their signs in the P-phase (Fig.~{}\ref{fig5}a). Note that the $S_1$ sign inversion occurs at the concentration of $x_S\sim0.05$ much lower than that of  $R_H$ ($x_R\approx0.115$, Fig.~{}\ref{fig5},a). The boundary dividing the $S_1>0$ and $S_1<0$ regions is very close to the crossover between classical and quantum critical fluctuations (line 2 in Fig.~{}\ref{fig1}) predicted for Mn$_{1-x}$Fe$_x$Si \cite{Demishev13}.  This crossover induced by the hidden QC point is also detected from resistivity data, which are sensitive to the scattering regime of charge carriers  \cite{Demishev13}. Therefore it is possible to suppose that the change of magnetic scattering on spin fluctuations taking place at the crossover line affects skew-scattering contribution to AHE  and may be responsible for its observed inversion in Mn$_{1-x}$Fe$_x$Si. However, no relevant theory describing this effect is available at this moment.

The other important finding appears from the OHE sign inversion detected at Fe content $x_R\approx0.115$ (Fig.~{}\ref{fig5},a). This fact points to the competing electron and hole contributions to charge transport in Mn$_{1-x}$Fe$_x$Si. Note that the boundary between negative $(R_H<0, x<x_R)$ and positive $(R_H>0, x>x_R)$ OHE coincides with the hidden QCP $x=x^*\approx0.11$ (Fig.~{}\ref{fig1},~line 1). Such an inversion of OHE may result from the change of the FS topology in strong magnetic fields, when the inversed cyclotron frequency is much less than the electrons $\tau_e$ and holes $\tau_h$ relaxation times. In this case, the corresponding Lifshitz transition at QCP would favor scenario based on Heisenberg-type model of magnetism \cite{Paschen04}. 

However, this QC scenario does not probably meet the case of Mn$_{1-x}$Fe$_x$Si. Firstly, the data in Fig.~{}\ref{fig3} correspond evidently to weak magnetic fields. Considering the effective concentrations $n(x)$ and $p(x)$ and mobilities $\mu_e(x)$ and $\mu_h(x)$ for electrons and holes, respectively, the condition for OHE sign inversion in the degenerate limit $p(x_R)-b(x_R)^2n(x_R)=0$  \cite{Seeger73} depends on the mobilities’ ratio $b(x)=\left|\mu_e(x)/\mu_h(x)\right| $. 
Secondly, high electron effective mass $m_e\sim17m_0$ \cite{Mena03} makes difficult reaching the high magnetic field regime. Therefore the observed $R_H(x)$ sign inversion  does not exclude itinerant origin of Hall effect peculiarities \cite{Combier13}. At the same time the OHE evolution in Mn$_{1-x}$Fe$_x$Si suggests that the substitution of Mn with Fe results in the effective hole doping so that the FS definitely evolves. This opportunity is not foreseen in the itinerant models of QC phenomena in MnSi based solids  \cite{Tewari06,Kruger12}.   

Quantitative information about FS evolution crucial for the analysis of the QC phenomena in Mn$_{1-x}$Fe$_x$Si can be extracted from our $R_H(x)$ data under some model assumptions. Firstly, the main effect of substitution of Mn by Fe is suggested to be the change of electron $n$ and hole $p$ concentrations. This hypothesis based on the experimental data (Fig.~{}\ref{fig5}a) means that electrons are associated with manganese whereas holes are supplied by iron. The simplest case assumes $n(x)=n(0)(1-x)$ and $p(x)=p_1x$ ($p_1$ is some coefficient). Secondly, the  $b(x)=\left|\mu_e(x)/\mu_h(x)\right|$ ratio is treated as a constant for the studied concentrations. The latter supposition to be apparently rather rough approximation is argued below.

For $b(x)=$const the expression for two groups of charge carriers  \cite{Seeger73} may be reduced to 
\begin{equation}
R_H(x)=R_H(0)\dfrac{1-x/x_R}{(1+a\cdot{x}/x_R)^2},
\end{equation}
with $R_H(0)=-(n(0)\left|e\right|)^{-1}$  and $a=b(1-x_R)-x_R$. $R_H(x_R)=0$ fixes the value of $p_1=n(0)b_2(1-x_R)/x_R$. For $x_R\approx0.115$ two parameter fitting by Eq.~(1) describes reasonably the $R_H(x)$ data with $R_H(0)=–(1.48\pm0.16)\cdot10^{-4}$~cm$^3$/C and $a=0.13\pm0.04$ $ (b\approx0.28)$ (solid line in Fig.~{}\ref{fig5}a). The nice correlation of the fit with the $R_H(x)$ points (Fig.~{}\ref{fig5}a) proves our supposition for $b(x)=$const.

	Another feature can be captured from the concentration dependence of conductivity $\sigma(x)$ in Mn$_{1-x}$Fe$_x$Si. Allowing for $b=m_h\tau_e/m_e\tau_h=$const $(m_{e,h}$ and $\tau_{e,h}$ are effective masses and relaxation times for electrons and holes) the total conductivity may be expressed as
	\begin{equation}
\sigma(x)=\sigma(0)(1+\gamma\cdot{x})\tau_e(x)/\tau_e(0),
	\end{equation}
where $\gamma=b(1/x_R-1)-1\approx1.13$. For $\tau_e(x)=$const conductivity is expected to increase linearly with $x$ (line~4 in Fig.~{}\ref{fig5}b). So a pronounced decrease of the $\sigma(x)/\sigma(0)$  ratio found at $x<0.3$ for $T=30$~K (squares in Fig.~{}\ref{fig5}b) suggests the strong concentration dependence of relaxation time as estimated from Eq.~(2) (circles in Fig.~{}\ref{fig5}b). As scattering on spin fluctuations dominates in Mn$_{1-x}$Fe$_x$Si  \cite{Demishev13,Demishev12,Demishev14}, it is reasonable to suppose that this mechanism affects equally on electrons and holes providing constant ratio of their mobilities and relaxation times. Because spin fluctuations control also the electron spin resonance linewidth $W(x)$ in Mn$_{1-x}$Fe$_x$Si \cite{Demishev14}, $\tau_{e,h}\sim1/W(x)$ is expected in the considered model. The dataset of $\tau_S(x)/\tau_S(0)\sim W(0)/W(x)$  \cite{Demishev14,Mena03} (stars in Fig.~{}\ref{fig5}b) demonstrates evident correlation between $\tau_S(x)$ and $\tau(x)$ behavior (Fig.~{}\ref{fig5}b) that can be considered as an additional justification for the suggested model.

	In our results iron doping fills the hole FS pocket and Mn$_{1-x}$Fe$_x$Si becomes a semimetal at intermediate Fe content. In pure MnSi holes are missing so that Lifshitz transition occurs formally at $x=0$ not relating to the hidden QCP $x^*$. Besides, a smooth filling of the electron and hole FS pockets is expected in the studied range $x<0.3$. The results need to be checked by rigorous band calculations and ARPES experiments.

	Our insight on Mn$_{1-x}$Fe$_x$Si allows suggesting some verifiable implications on the hierarchy of magnetic interactions in this system.  Despite the fact that these compounds are usually treated as itinerant magnets   \cite{Grigoriev09,Grigoriev11,Nishihara84,Bauer10,Griffiths69,Tewari06,Kruger12} few works discuss microscopic origin of QC phenomena in these solids in itinerant paradigm. According to \cite{Kruger12}, the \textquotedblleft driving force\textquotedblright for quantum phase transitions is the variation of Hubbard repulsion energy, possible effects of FS transformation being not allowing for. In our opinion, these effects should not be disregarded in itinerant approach even if they can not be easily incorporated into the existing models \cite{Tewari06,Kruger12}.

	Heisenberg picture of magnetism in MnSi and Mn$_{1-x}$Fe$_x$Si suggested from magnetic resonance  \cite{Demishev12,Demishev14,Demishev11}, magnetoresistance  data \cite{Demishev12} and the LDA calculations  \cite{Corti07} predicts the modulation of exchange interactions under substitution of Mn with Fe. Generally, magnetic interactions in $3d$-metals are not described by RKKY approach due to itinerant origin of magnetism \cite{Vonsovskii}. However, if LMMs of Mn are either postulated or deduced for Mn$_{1-x}$Fe$_x$Si from experiment and theory \cite{Corti07,Demishev12,Demishev14,Demishev11}, they have to interact via the RKKY exchange. In this case, the exchange energies can be estimated assuming quadratic isotropic dispersion laws for electrons and holes. Extending of the general expression for the RKKY interaction  \cite{Vonsovskii} to the two groups of charge carriers results in 
	\begin{equation}
	J(x,r)=J(0,r)\dfrac{\varphi(\alpha(r)(1-x)^{1/3})+(m_h/m_e)\varphi(\alpha^\prime(r)x^{1/3})}
	{\varphi(\alpha(r))}  
	\end{equation}
with $\varphi(z)=z\cos(z)-\sin(z)$, $\alpha(r)=2k_{Fe}(0)r$  and $\alpha^\prime(r)=\alpha(r)(a+x_R)^2/\left[ x_R\left( 1-x_R\right) \right] $ (see \cite{Sup1} for details).

\begin{figure}[t]
\includegraphics[width=0.7\linewidth]{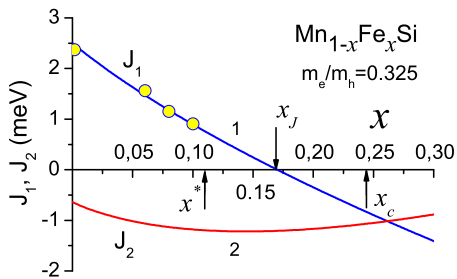}
\caption{\label{fig6}(Colour on-line) The estimated $nn$ $J_1(r_1)$ and $nnn$ $J_2(r_2)$ exchange constants in Mn$_{1-x}$Fe$_x$Si ($m_e/m_h=0.325$). Circles represent the experimental $J(x)$ data  \cite{Grigoriev11}}
\end{figure}

Eq.~(3) allows estimating magnetic interactions in MnSi and comparing them with available data \cite{Grigoriev09}. The nearest neighbor $(nn)$ and next nearest neighbor $(nnn)$ distances between Mn sites in the B20 cubic structure are $r_1=2.80$~\AA{} and $r_2=4.39$~\AA{}, respectively  \cite{Grigoriev10}. Allowing for the experimental values of $nn$ exchange $J_1=J(r_1)\approx2.5$~meV \cite{Grigoriev11} and the calculated Fermi wave vector $k_{Fe}(0)\approx1.08\cdot10^8$ cm$^{-1}$, the RKKY function $J(0,r)$ gives $J_2=J(r_2)\approx$--0.6~meV (Fig.~{}\ref{fig6}). Thus $nn$ and $nnn$ exchanges in pure MnSi  prove to be F $(J_1>0)$ and antiferromagnetic (AF) $(J_2<0)$ ones. 

Fitting of the experimental $J(x)$ data for Mn$_{1-x}$Fe$_x$Si \cite{Grigoriev11} by Eq.~(3) shows that the hole contribution (the second term in the nominator) is important as long as it stays negative and its absolute value increases noticeably with $x$. At the same time the electron term in the nominator does not vary significantly not accounting for the $J(x)$ evolution (circles in Fig.~\ref{fig6}b). The best fit of the $J_1(x)$ data within Eq.~(3) (line~1 in Fig.~{}\ref{fig6}b) corresponds to $m_e/m_h=0.325$. If $m_e/m_h$ is fixed Eq.~(3) gives the $nnn$ exchange $J_2(x)=J(x,r_2)$ (line~2 in Fig.~{}\ref{fig6}b) without any additional parameters.

	Several important findings can be established from the $J_1(x)$ and $J_2(x)$ dependences. First of all, the $nn$ exchange changes sign at $x_J\sim0.17$ (Fig.~{}\ref{fig6}b). Thus the FS evolution in Mn$_{1-x}$Fe$_x$Si induces a qualitative change of magnetic interaction from F $J_1(x<x_J)>0$ to AF $J_1(x>x_J)<0$. Secondly, the opposite signs of $J_1$ and $J_2$ for $x<x_J$ and the same signs of $J_1$ and $J_2$ for $x>x_J$ mean the strong influence of frustration   on the magnetic properties of Mn$_{1-x}$Fe$_x$Si pointed out earlier for MnSi based solids \cite{Hopkinson07}. Because the absolute values of $J_1$ and $J_2$ become equal near QCPs $x^*$ and $x_c$, frustration should essentially affect resulting spin configuration. For example, the discrepancy between the LRO temperature $T_c(x)$ and exchange energy $J(x)$ \cite{Friedemann10} (Fig.~{}\ref{fig1}) may be induced by $nnn$ AF interaction. Moreover, the frustration effects are expected to be strong in the \textquotedblleft tail\textquotedblright of SRO phase for $x>x^*$ (Fig.~{}\ref{fig1})  facilitating segmentation into spin clusters in the G phase ($x>x_c$) \cite{Demishev13}. This suggestion resolves the abovementioned paradox  that the percolation for Mn magnetic ions is not formally broken under the topological transition in Mn$_{1-x}$Fe$_x$Si \cite{Demishev13}.

In summary, the novel approach to Hall effect study in the P phase of Mn$_{1-x}$Fe$_x$Si allowed finding the dependences of the OHE and AHE constants $R_H$ and $S_1$ on the Fe content. Hole doping induced by the substitution of Mn with Fe proves that two groups of charge carriers contribute to OHE in Mn$_{1-x}$Fe$_x$Si. The fact that the observed $R(x)$ and $S_1(x)$ sign inversions are definitely associated with the hidden QCP $x^*\sim0.11$  reveals the relationship of these transport anomalies to the QC transition between LRO and SRO phases.

	Our quantitative analysis leads to some predictions in the field of fermiology, magnetic interactions and QC phenomena in Mn$_{1-x}$Fe$_x$Si to be verified by experiments. In particular, rising of Fe content is expected to reduce the electron FS section filling the  pocket with heavier holes ($m_h/m_e\approx3$). The discovered semimetallic behavior is not foreseen by the itinerant models of quantum criticality in Mn$_{1-x}$Fe$_x$Si. In contrary, the LMM approach predicts strong frustration affecting the position of the hidden QCP $x^*\sim0.11$ and facilitating (together with the disorder) the SRO suppression and the formation of the G phase for $x>x_c\sim0.24$. As long as the exchange energies are tuned via the RKKY mechanism, the change of electron and hole concentrations may be considered as a microscopic \textquotedblleft driving force\textquotedblright for QC in Mn$_{1-x}$Fe$_x$Si.

	The most striking consequence of the LMM approach is the F to AF $J_1(x)$ evolution    occurring in Mn$_{1-x}$Fe$_x$Si at $x_J\sim0.17$. The different types of magnetic interactions modulated by Fe content make the SRO phase to be inhomogeneous in the range $x^*<x<x_c$. Simultaneously, the G phase $(x>x_c$) should be constructed from AF rather than F spin clusters. Besides, the DMI and the spiral structures need to be treated by different ways for $x>x_J$ and $x<x_J$. As far as this unusual behaviour can be hardly expected in itinerant models  \cite{Tewari06,Kruger12}, verifying of this prediction may be considered as an \textit{experimentum crucius} for Heisenberg-type models of magnetism suggested for Mn$_{1-x}$Fe$_x$Si.

	The authors are grateful to D.I.Khomskii and N.E.Sluchanko for stimulating discussions. This work was supported by the RAS Programmes \textquotedblleft Electron spin resonance, spin-dependent electronic effects and spin technologies\textquotedblright, \textquotedblleft Electron correlations in strongly interacting systems\textquotedblright and by RFBR grant 13-02-00160.

\bibliography{HalleffectMnFeSi}
 
\end{document}